\documentclass[12pt]{iopart}

\usepackage{epsfig}
\begin{document}

\title{Quark-hadron mixed phases in protoneutron stars}

\author{Giuseppe Pagliara, Matthias Hempel and J\"urgen Schaffner-Bielich}

\address{Institut f\"{u}r Theoretische Physik, Ruprecht-Karls-Universit\"at,
   Philosophenweg 16,  D-69120, Heidelberg, Germany}
\ead{pagliara@thphys.uni-heidelberg.de}

\begin{abstract}
We consider the possible formation of the quark hadron mixed phase in
protoneutron stars. We discuss two cases: the first one, corresponding
to a vanishingly small value of the surface tension of quark matter,
is the well known mixed phase in which the global electric charge
neutrality condition is imposed. In turn, this produces a non-constant
pressure mixed phase. In the second case, corresponding to very large
values of the surface tension, the charge neutrality condition holds
only locally. However, the existence in protoneutron star matter of
an additional globally conserved charge, the lepton number, allows for
a new type of non-constant pressure mixed phase. We discuss the
properties of the new mixed phase and the possible effects of its
formation during the evolution of protoneutron stars.
\end{abstract}

\maketitle

\section{Introduction}
The possibility of a first order phase transition from nuclear matter
to quark matter at the high densities and moderate temperatures
reached in Supernova events or, afterwards, in the still hot and
lepton rich protoneutron stars, has attracted much attention in the
last years
\cite{Prakash:1995uw,Drago:1997tn,Steiner:2000bi,Panda:2003dh,Nicotra:2006eg,Pons:2001ar,Nakazato:2008su,Sagert:2008ka,Drago:2008tb,Yasutake:2009kj,Bombaci:2009jt,Mintz:2009ay}. In
particular, possible interesting fingerprints of this phase transition
within the neutrino signal emitted from the newly born neutron star,
have been proposed in
\cite{Pons:2001ar,Yasutake:2009kj,Sagert:2008ka,Dasgupta:2009yj} thus
providing a new tool to investigate the properties of the QCD phase
diagram at high densities.

The usual way to model the phase transition, after the seminal
paper of Glendenning \cite{Glendenning:1992vb}, is to impose the Gibbs
conditions for a first order phase transition in muticomponent
systems. For matter in neutron stars indeed the baryonic charge and the
electric charge are conserved globally resulting in an extended mixed
phase in which the pressure increases as a function of the baryon density. 
In turn, this implies the possible existence of a layer of mixed phase 
embedded between the pure nuclear and pure quark
phases in hybrid stars. Within the mixed phase, due to the finite value of the
surface tension between hadronic matter and quark matter, different
structures appear depending on the density: first the quark pasta
phases, immersed in nuclear matter, and then the nuclear pasta
phases immersed in quark matter
\cite{Heiselberg:1992dx,Glendenning:1995rd}. However, as shown in
Refs.~\cite{Voskresensky:2002hu,Endo:2005zt}, sophisticated calculations
which include finite size and charge screening effects together with
more recent estimates for the value of the surface tension, have shown that the mixed phase
obtained with the Gibbs construction is actually quite similar to the
simpler Maxwell construction. Due to the large value of the surface
tension, the charged structures have sizes larger than the electron
Debye screening length and therefore their net charge is significantly
reduced. One is back to the (almost) constant pressure Maxwell
construction which implies that the mixed phase cannot be present in
neutron stars where the pressure must be a monotonic function of the
radius.

In protoneutron stars, however, an additional charge is conserved,
the lepton number, during the stage of neutrino trapping. As we will 
show in the following, due to this additional conservation law,
it is possible to form a new type of mixed phase in protoneutron stars.
Since this mixed phase is related to the conservation of lepton number,
this phase will gradually ``disappear'' as the neutrinos become untrapped.

\section{The equation of state of mixed phases}
We want to present here two extreme cases for the mixed phase in
protoneutron star matter: the first, widely considered in the
literature, corresponding to a vanishing surface tension (we call it
Mixed Phase 1, MP1) and the second, corresponding to a
very large value of the surface tension (we call it Mixed Phase 2,
MP2). Let us discuss first the structure of the interface between
hadronic matter and quark matter: a charge separated interface is
formed with a size of the order of the Debye screening length, $\sim
10$ fm, with a layer of positively charged, electron depleted,
hadronic matter on one side and a layer of quark matter with an excess
of the electrons on the other side (as discussed in
\cite{Alford:2001zr} for the CFL phase). The interface is stabilized
by the resulting electric field. Notice that neutrinos, being not
affected by the electric field, can freely stream across the
interface. Consequently, the lepton number is conserved globally even
in the case of a large surface tension.

We consider the ``standard'' conditions of ProtoNeutron Star (PNS) matter
\cite{Steiner:2000bi}: fixed lepton fraction
$Y_L=(n_e+n_{\nu})/n_B=0.4$ and fixed entropy per baryon $S/N_B=1$
where $n_e$, $ n_{\nu}$ and $n_B$ are the electron, neutrino and
baryon number densities and $S$ is the entropy.  The chemical
equilibrium between the different species of particles within the two
phases allows to write the following general relations:

\begin{eqnarray}
\mu_n=\mu_B^H,\,\, \mu_p = \mu_B^H+\mu_C^H,\,\,  \mu_e^H=\mu_L^H-\mu_C^H,\,\, \mu_{\nu}^H=\mu_L^H\\
\mu_u=\frac{\mu_B^Q+ 2\mu_C^Q}{3},\,\,  \mu_d=\frac{\mu_B^Q-\mu_C^Q}{3},\,\, \mu_s=\frac{\mu_B^Q-\mu_C^Q}{3},\nonumber\\ 
\mu_e^Q= \mu_L^Q-\mu_C^Q,\,\, \mu_{\nu}^Q=\mu_L^Q
\end{eqnarray}

where $\mu_i^A$ ($i=n,p,e,\nu,u,d,s$ and $A=H,Q$) are the chemical potentials
of neutrons, protons, electrons, neutrinos and up, down, strange quarks 
within the hadronic phase (A=H) and the quark phase (A=Q).  
The chemical potentials associated with globally conserved quantities are
equal in the two phases \cite{Hempel:2009vp}. In presence of a vanishingly small value of
the surface tension, the electric charge is conserved globally as the baryonic number and the lepton number,
therefore its chemical potential is continuous at the onsets of the phase
transition and we can write the Gibbs conditions as follows:

\begin{eqnarray}
P^H(\mu_B^H,\mu_C^H,\mu_L^H,T^H) = P^Q(\mu_B^Q,\mu_C^Q,\mu_L^Q,T^Q)\\
(1-\chi)n^H_C + \chi n^Q_C -  n_e=0\\
n_e+n_{\nu}= Y_L n_B\\
(1-\chi)s^H+\chi s^Q= S/Nn_B\\
T^H=T^Q\\
\mu_B^H=\mu_B^Q,\,\, \mu_C^H=\mu_C^Q,\,\,\mu_L^H=\mu_L^Q
\end{eqnarray}

where $P^{H,Q}$, $T^{H,Q}$, $s^{H,Q}$ and $n_C^{H,Q}$ 
are the pressure, the temperature, the entropy density and the electric charge density of the hadronic and quark phases 
respectively. $\chi$ is the volume fraction of the quark phase.

In presence of a large value of
the surface tension, the electric charge is conserved only locally,
therefore its chemical potential is different in the two phases,
but the lepton number, as noticed before, is still conserved globally as the baryon number
and its chemical potential must be continuous across the phase transition.
In this case, we can write the Gibbs conditions as follows:

\begin{eqnarray}
P^H(\mu_B^H,\mu_C^H,\mu_L^H,T^H) = P^Q(\mu_B^Q,\mu_C^Q,\mu_L^Q,T^Q) \\
n^H_C-n^H_e=0\\
n^Q_C-n^Q_e=0\\
(1-\chi)(n^H_e+n_{\nu})+ \chi(n^Q_e+n_{\nu})= Y_L n_B\\
(1-\chi)s^H+\chi s^Q= S/N n_B\\
T^H=T^Q\\
\mu_B^H=\mu_B^Q,\,\,\mu_L^H=\mu_L^Q
\end{eqnarray}

\begin{figure}
\vskip 0.7cm
    \begin{centering}
\epsfig{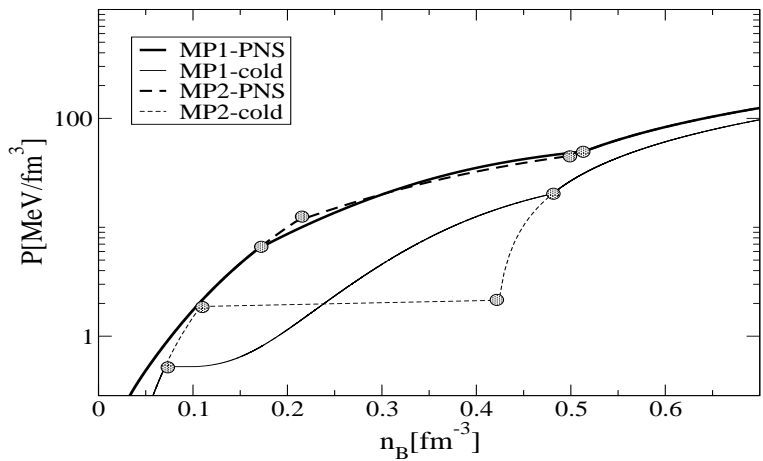}
    \caption{The equations of state for protoneutron star matter and
cold and beta stable matter for the two types of mixed phases MP1 and MP2. The dots
indicate the onset and the end of the mixed phases. The mixed phases 
for protoneutron star matter MP1 and MP2 are quite similar. For 
cold and beta stable matter, MP2 coincides with a constant pressure
Maxwell construction. 
\label{fig1} }
   \end{centering}
   \end{figure} 

The systems of equations above presented have a general validity. To
give some numerical examples, we consider, as customary, two models for
strongly interacting matter: one at low density with nucleon degrees
of freedom and one at large densities with quark degrees of freedom.
For the former, we adopt the relativistic mean field model with the
parameterization TM1 \cite{Shen:1998gq} and for the
latter the MIT bag model. We set the masses of up and down quarks to
zero and the mass of the strange quark to $100$ MeV. The bag constant
$B$ is set to $B^{1/4}=165$ MeV. The phase
transition is then computed by using the systems for the two mixed phases MP1 and MP2. In
Fig.~\ref{fig1}, we show the equations of state for protoneutron star
matter and cold and beta stable matter for the two cases MP1 and MP2.
Notice that for PNS matter, the equations of state are quite similar:
an extended mixed phase with varying pressure is obtained also in the
case of local charge neutrality due to the existence of an additional
globally conserved quantities which is the lepton number. In cold and
beta stable matter on the other hand, there are only two conserved
charged, the baryonic and the electric charge, and therefore in the
case of MP1 one obtains an extended mixed phase but in MP2
the result is the simple constant pressure Maxwell construction.
In general, the mixed phase MP2 extends over a smaller range of density and 
has a lower pressure gradient with respect to MP1, because the requirement of local charge
neutrality is more restrictive.

\section{Discussion and conclusions}
\begin{figure}
\vskip 0.7cm
    \begin{centering}
\epsfig{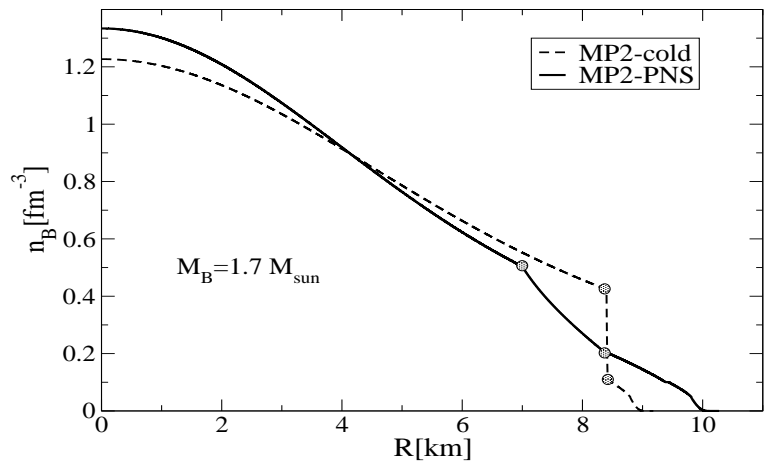}
    \caption{Density profiles for a $M_B=1.7 M_{sun}$ star in its 
protoneutron star stage and the cold and beta stable asymptotic state, the case
of the mixed phase MP2 is considered. In the protoneutron star a $2$km layer of mixed phase
is present while in the cold configuration the mixed phase disappears and 
a sharp interface separating the pure quark and nuclear phase is obtained. 
\label{fig2} }
   \end{centering}
   \end{figure}

The new mixed phase here proposed, MP2, is 
present as long as neutrinos are trapped. After the complete deleptonization and cooling of the star
it becomes a constant pressure mixed phase which cannot 
appear in neutron stars. This implies that the deleptonization drives
a gradual modification of the structure of the star: for sufficiently large
masses of the protoneutron stars, one has a core of pure quark matter, a layer of 
mixed phase MP2, and then a layer/crust of nuclear matter.
The layer of mixed phase disappears during deleptonization and, 
at the end of the deleptonization stage, a sharp interface will 
separate the pure quark and the pure nuclear phase, see Fig.~\ref{fig2}.

Notice that the mixed phase MP2 has quite different properties with
respect to MP1: in MP2 the two phases are locally charge neutral,
therefore no Coulomb lattice with charged finite structures of the two
phases can form. This implies also that only spherical pasta
structures can be formed since the 1-D and 2-D structures can take
place only in the presence of Coulomb interactions. Moreover the charge
neutral structures have macroscopic sizes contrary to the MP1 case
where the optimal size of the structures is limited by the
Coulomb energy. We expect that within the MP2 mixed phase
the process of diffusion of neutrinos can be significantly different
with respect to the MP1 case: for instance, due the large sizes of the
structures within MP2, no coherent scattering of neutrinos with pasta
structures can take place \cite{Reddy:1999ad}, as the neutrino
wavelength is of the order of tens of fermi. A detailed simulation of
neutrino transport within this new mixed phase would be extremely
interesting for the possible implications on the neutrino signal of
the changes of the structure of the star during deleptonization.  Also
the motion and the interactions of the drops/bubbles within the mixed
phase, in presence of turbulence, might represent an interesting
source of gravitational waves. 
A more detailed
study, which includes also superconducting quark matter \cite{Ruester:2005ib,Sandin:2007zr,Pagliara:2007ph}, 
is presently in progress.
\\\\
{\bf Acknowledgments:}

The work of G.~P. is supported by the Deutsche Forschungsgemeinschaft
(DFG) under Grant No. PA 1780/2-1. M.~H. acknowledges support 
from the Graduate Program for Hadron and Ion Research (GP-HIR). 
J.~S.~B. is supported by the DFG
through the Heidelberg Graduate School of Fundamental Physics. This work
was also supported by CompStar, a Research Networking Programme of the
European Science Foundation.

\end{document}